\documentclass[prl,preprint,showpacs]{revtex4}
\usepackage{bm}
\begin{document}
\title{Magnetization of nanoparticle systems in a rotating magnetic field}
\author{S.~I.~Denisov,$^{1,2}$ T.~V.~Lyutyy,$^{2}$ and
P.~H\"{a}nggi$^{1}$}
%\email{stanislav.denisov@physic.uni-augsburg.de}
\affiliation{$^1$Institut f\"{u}r Physik, Universit\"{a}t
Augsburg, Universit\"{a}tsstra{\ss}e 1, D-86135 Augsburg, Germany\\
$^2$Sumy State University, 2 Rimsky-Korsakov Street, 40007 Sumy,
Ukraine}

%\date{submitted to Physical Review Letter: \today}

\begin{abstract}
The investigation of a sizable thermal enhancement of magnetization is put
forward for uniaxial ferromagnetic nanoparticles that are placed in a  rotating
magnetic field.  We elucidate the nature of this phenomenon and evaluate the
resonant frequency dependence of the induced  magnetization. Moreover, we
reveal the role of  magnetic dipolar interactions, point out potential
applications and reason the feasibility of an experimental observation of this
effect.
\end{abstract}
\maketitle

Presently, the study of magnetic nanoparticles and their structures is one of
the most important research areas in nanoscale physics. A first reason is that
such nanoparticles increasingly find numerous applications that range from
medicine to nanotechnology. Another reason is that these systems exhibit a
number of remarkable physical phenomena, such as quantum tunneling of
magnetization \cite{tunneling}, giant magnetoresistance \cite{m_resistance},
exchange bias \cite{exch_bias}, and finite-size and surface effects
\cite{size_effects}, to name but a few. Moreover, the study of fundamentals of
magnetic behavior in these systems is also an important issue, especially for
high-density data storage devices \cite{nanostruc}.

From a practical point of view, the lifetime of stored data and the switching
time (i.e., the time during which the reversal of the nanoparticle magnetic
moments occurs) are salient characteristics of such devices. Now, thanks to the
experimental discovery of fast switching of magnetization \cite{switching}, the
switching time reaches the fundamental (picosecond) limit for field-induced
magnetization reversal. On the contrary, a feasible lifetime must cover up to
ten years and beyond. Its value is usually limited by the superparamagnetic
effect \cite{Neel} and is defined by the probabilities $p_{\sigma}$ that the
nanoparticle magnetic moment $\mathbf{m}$ stays in the up ($\sigma = +1$) and
down ($\sigma = -1$) equilibrium directions. These probabilities, which are
also responsible for other thermal effects in such systems including magnetic
relaxation \cite{DFT}, are very sensitive to small perturbations that change
the \textit{static} states of the magnetic moments. Namely, according to the
Arrhenius law \cite{HTB} the ratio $p_{+1}/p_{-1}$ is approximately given by
$\exp(\Delta E/kT)$, where $\Delta E = E_{+1} - E_{-1}$, $E_{\sigma}$ is the
potential barrier for the reorientation $\sigma \to -\sigma$, $k$ is the
Boltzmann constant, and $T$ is the absolute temperature. Therefore, if without
perturbations $\Delta E = 0$, then $p_{+1}/ p_{-1} = 1$ and the nanoparticle
system is demagnetized. But due to the exponential dependence on $\Delta E$ and
$T$, the ratio $p_{+1}/p_{-1}$ can drastically be changed by small
perturbations. In particular, a static magnetic field $H$ applied along the
nanoparticle easy axis of magnetization yields $\Delta E = 2Hm$ ($m =
|\mathbf{m}|$), and so $p_{+1}$ strongly differs from $p_{-1}$ if $|H|/H_{a}
\gg 1/4a$ where $a = H_{a}m/2kT$, and $H_{a}$ is the anisotropy field. This
means that even small magnetic fields (in comparison with $H_{a}$) almost fully
magnetize the nanoparticle systems when $a \gg 1$.

In the case of time-periodic perturbations the situation is not settled yet and
far less researched. On the one hand, these perturbations generate
\textit{dynamical} states of the nanoparticle magnetic moments that, because of
their natural precession, are expected to be different for the up and down
magnetic moments. Therefore, a dynamical magnetization of the system, i.e.,  a
magnetization which is induced by periodic perturbations at $T=0$, can exist
but it is expected to be small and, at first sight, it cannot be changed by
thermal fluctuations since these perturbations do not change the mentioned
above potential barriers. On the other hand, it may be expected that, due to
the different precessional states of the up and down magnetic moments, the
probabilities $p_{\sigma}$ are different and thus thermal fluctuations
contribute to the dynamical magnetization. In this Letter, we attempt to solve
this challenge which comprises both basic and applied aspects of the concept of
a mean first-passage time.

\textit{Dynamical magnetization.}---To calculate the dynamical magnetization,
we consider the case of identical, non-interacting nanoparticles whose easy
axes of magnetization are parallel to each other and the dynamics of the
magnetic moment $\mathbf{m}$ is governed by the deterministic Landau-Lifshitz
equation \cite{LL}
\begin{equation}
    \dot\mathbf{m} = -\gamma\mathbf{m}\times\mathbf{H}_{eff}
    -\frac{\lambda\gamma}{m}\,\mathbf{m}\times
    (\mathbf{m}\times\mathbf{H}_{eff}).
    \label{eq:L-L}
\end{equation}
Here $\gamma(>0)$ is the gyromagnetic ratio, $\lambda(>0)$ is the dimensionless
damping parameter, $\mathbf{H}_{eff} = -\partial W/\partial \mathbf{m}$ is the
effective magnetic field acting on $\mathbf{m}$, and $W$ is the nanoparticle
magnetic energy. If the easy exes are parallel to the $z$ axis and the external
magnetic field $\mathbf{h} (t)$ is circularly polarized in the $xy$ plane,
i.e., $\mathbf{h}(t) = h(\cos \omega t, \rho\sin\omega t,0)$, where $h =
|\mathbf{h} (t)|$, $\omega$ is the frequency of field rotation, and $\rho = -1$
or $+1$ (the sign $-$ corresponds to the clockwise rotation of $\mathbf{h}(t)$
and the sign $+$ to the counterclockwise one), then
\begin{equation}
    \textstyle W = {1\over 2} mH_{a}\sin^{2}\theta - mh\sin\theta\cos\psi
    \label{W}
\end{equation}
($\psi = \varphi - \rho \omega t$, and $\theta$ and $\varphi$ are the polar and
azimuthal angles of $\mathbf{m}$, respectively). For this case, the solution of
Eq.~(\ref{eq:L-L}) is well studied in the context of ferromagnetic resonance
\cite{GM} and nonlinear magnetization dynamics \cite{BSM}. Specifically, the
precession angles $\theta_ {\sigma}$ of the up and down magnetic moments (see
Fig.~1) at $\theta_ {\sigma} \ll 1$ are given by
\begin{equation}
    \theta_{\sigma} = \frac{(1 + \lambda^{2})\gamma h}{\sqrt{[(1
    + \lambda^{2})\omega_{r} - \rho\sigma\omega]^{2} +
    \lambda^{2}\omega^{2}}},
    \label{theta_sigma}
\end{equation}
where $\omega_{r} = \gamma H_{a}$. Note that the angles $\theta_
{\sigma}$ exhibit the resonance dependence on $\omega$ only for
$\rho\sigma = +1$.

We define the dimensionless magnetization of the nano\-particle system induced
by the magnetic field $\mathbf{h}(t)$ as $\mu =(1/N)\sum_{i=1}^{N} m_{zi}/m$
(the index $i$ labels the nanoparticles, $N \gg 1$). If the magnetic anisotropy
barrier essentially exceeds the thermal energy, i.e., $a \gg 1$, this
definition yields $\mu =(1/N)\sum_{\sigma} \sigma N_{\sigma}\cos \theta_
{\sigma}$ ($N_{\sigma}$ is the number of magnetic moments in the state
$\sigma$, $N_{-1} + N_{+1} = N$), and so $\mu =\sum_{\sigma}\sigma p_{\sigma}
\cos \theta_ {\sigma}$ ($p_{\sigma} = N_{\sigma}/N$). Assuming $p_{\sigma} =
1/2$ at $T = 0$ (this means that the nanoparticle system is demagnetized if $h
= 0$) and using the condition $\theta_ {\sigma} \ll 1$ and Eq.~(\ref
{theta_sigma}), for the dynamical magnetization $\mu_{d} = \mu|_{T=0} =
(\theta^{2} _{-1} - \theta^{2}_ {+1})/4$ we obtain
\begin{equation}
    \mu_{d} = -\rho \frac{(1+\lambda^{2})\gamma^{2}h^{2}
    \omega\omega_{r}}{[(1+\lambda^{2})\omega_{r}^{2} +
    \omega^{2}]^{2} - 4\omega^{2}\omega^{2}_{r}}.
    \label{eq:mu_d}
\end{equation}

According to this result, the direction of dynamical magnetization $\mu_{d}$
and the direction of magnetic field rotation follow the left-hand rule (the
reason is that the natural precession of the magnetic moments is
counterclockwise), and the dependence of $\mu_{d}$ on $\omega$ always has a
resonant character with $\max \mu_{d} = \mu_{d}|_ {\omega = \omega_{m}} $,
where $\omega_{m} = (\omega_{r}/\sqrt{3})[1 - \lambda^{2} + 2(1 + \lambda^{2} +
\lambda^{4})^{1/2}]^{1/2}$. But the value of $\mu_{d}|_ {\omega = \omega_{m}}$
is very small, however, because with $\max{\theta_{\sigma}} = h/\lambda H_{a}$
even for $\lambda \ll 1$, Eq.~(\ref{eq:mu_d}) yields $\mu_{d}|_ {\omega =
\omega_{m}} = -\rho (h/2\lambda H_{a})^{2}$. Note also that after switching on
the magnetic field $\mathbf{h}(t)$ the initially demagnetized system reaches
the steady-state magnetization $\mu_{d}$ during a time interval of the order of
$t_{r} = 2/\lambda \omega_{r}$ (we recall that for $T = 0$ the states $\sigma$
of the magnetic moments are not changed with time and $p_{\sigma} = 1/2$).

\textit{Thermal enhancement of the dynamical magnetization.}---If $T \neq 0$
then the dynamics of the magnetic moments becomes stochastic. In this case, due
to thermal fluctuations, the magnetic moments can perform random transitions
from the one state $\sigma$ to the other $-\sigma$ and the probabilities
$p_{\sigma}$ can thus depend on $\mathbf{h}(t)$. But, in contrast to a static
magnetic field, a rotating field has no preferential direction and so it does
not impact $p_{\sigma}$ directly. Nevertheless, the probabilities $p_{+1}$ and
$p_{-1}$ must be different in the presence of $\mathbf{h}(t)$. The reason is
that if the mean times $t_{\sigma}$ which the magnetic moments reside in the
states $\sigma$ are much larger than the precession time $2\pi/\omega$, then
the up and down magnetic moments spend almost all time near the conic surfaces
with the cone angles $\theta_{+1}$ and $\theta_{-1}$, respectively. Since these
angles are different, see Eq.~(\ref{theta_sigma}), the times $t_{\sigma}$ must
be different as well. Accordingly, because in the steady state $p_{\sigma} =
t_{\sigma}/ (t_{+1} + t_{-1})$, we conclude that the probabilities $p_{\sigma}$
are also different and so thermal fluctuations in fact do contribute to the
induced magnetization $\mu$.

Using the conditions $\theta_{\sigma} \ll 1$ and $a \gg 1$, from the definition
of $\mu$ we obtain $\mu = \mu_{t} + \mu_{d}$ (we neglect the term $\mu_{t}
(\theta^{2} _{-1} + \theta^{2}_ {+1})/4$), where $\mu_{t} = p_{+1} - p_{-1}$ is
the desired contribution arising from the joint action of thermal fluctuations
and rotating field. According to the above argumentation, the condition
$p_{\sigma} < p_{-\sigma}$ holds if $\theta_{\sigma} > \theta_ {-\sigma}$. This
implies that the thermal contribution $\mu_{t}$ always enhances the
deterministic part $\mu_{d}$. Moreover, one expects the enhancement
\textit{increases} with \textit{decreasing} temperature. It is important to
note in this context that $\mu$ denotes the equilibrium magnetization which is
established during a time interval of the order of the transition time $t_{tr}$
between the states $\sigma$ and $-\sigma$ ($t_{tr} \sim \max{t_{\sigma}} \gg
t_{r}$ and $t_{tr} \to \infty$ as $T \to 0$). At these times the probabilities
$p_{+1}$ and $p_{-1}$ are generally different and thus $\lim_{T \to 0}\mu \neq
\mu_{d}$.

For determining the mean residence times $t_{\sigma}$ which define the
probabilities $p_{\sigma}$ and the magnetization $\mu_{t}$, we used the mean
first-passage time formalism \cite{HTB,HT}. Its application to our situation is
well-founded because in the case of small rotating field the magnetic moments
that are in the state $\sigma$ reach any point of the separatrix, which
separates the up and down states, with almost the same probability density.
Given that the stochastic dynamics of the magnetic moments is Markovian
\cite{Brown}, the standard mean first-passage procedure is employed in order to
account for the influence of the rotating field. Specifically, starting out
from the two-dimensional backward Fokker-Planck equation \cite{HT,DY} in the
rotating frame, we succeeded to derive a mathematically tractable and
physically transparent expression for the mean residence times:
\begin{equation}
    t_{\sigma} = t_{0} \exp[a(- \theta_{\sigma}^{2} +
    \sigma 2 H_{\text{eff}}/H_{a})],
    \label{t sigma}
\end{equation}
where $t_{0} = t_{r} \sqrt{\pi/4a} \exp a$ is the mean time which
the magnetic moment spends in the up or down state at $h = 0$, and
$H_{\text {eff}} = -\rho p \gamma h^{2}/\omega$ ($|H_{\text{eff}}|
\ll H_{a}$, $p \sim 1$).

As follows from (\ref{t sigma}), the rotating magnetic field influences the
mean residence times $t_{\sigma}$ via two  different mechanisms. A first one
consists in the appearance of the dynamical states of the magnetic moments
which are characterized by the precession angles $\theta _{\sigma}$. The
contribution of these states to $t_{\sigma}$ is governed by the first term in
the argument of the exponential function, which always decreases $t_{\sigma}$.
The second mechanism consists in changing the effective potential barrier
between the up and the down states. Its contribution to $t_ {\sigma}$ is
described by the second term in the argument of the exponential function, where
$H_{\text{eff}}$ can be interpreted as a \textit{static} effective magnetic
field applied \textit{along} the easy axis of magnetization. Since the sign of
$H_{\text{eff}}$ depends on $\rho$, this mechanism can either increase or
decrease $t_{\sigma}$.

Using Eqs.~(\ref{t sigma}) and (\ref{theta_sigma}), the definition $\mu_{t} =
(t_{+1} - t_{-1})/(t_{+1} + t_{-1})$ leads to our main result
\begin{equation}
    \mu_{t} = \tanh [2a(\mu_{d} + H_{\text{eff}}/H_{a})].
    \label{ind magn1}
\end{equation}
It shows that the rotating field always magnetizes the nanoparticle system
perpendicular to the plane of field rotation and the direction of magnetization
is uniquely defined by the direction of field rotation: $\textrm{sgn}\, \mu
_{\infty} = -\rho$. In general, both mentioned mechanisms contribute to
$\mu_{t}$. However, in the most interesting resonant case, when $\omega \sim
\omega_{m}$ and $\lambda \ll 1$, the first mechanism is dominating ($\mu_{d}
H_{a} / H_{\text{eff}} \sim \lambda^{-2} \gg 1$), and thus the magnetization
(\ref{ind magn1}) becomes
\begin{equation}
    \mu_{t} = \tanh (2a\mu_{d}).
    \label{ind magn2}
\end{equation}

According to this relation, being valid for $a \gg 1$ and $|\mu_{d}| \ll 1$,
the condition $\mu_{t}/ \mu_{d} \gg 1$ always holds (specifically, if $a
|\mu_{d}| \ll 1$ then $\mu_{t}/ \mu_{d} \approx 2a$). This means that a small
dynamical magnetization is strongly enhanced by thermal fluctuations, i.e.,
$\mu \approx \mu_{t}$. Comparing (\ref{ind magn2}) with the magnetization of an
Ising paramagnet, $\tanh(mH/kT)$, we see that the magnetic field rotating in a
plane perpendicular to the easy axes of the nanoparticles acts as a static
magnetic field $H = H_{a}\mu_{d}$, which is applied along these axes. As in the
case with $\mu_{d}$, the induced magnetization $\mu_{t}$ as a function of
$\omega$ exhibits a resonance character. The dependence of $\mu_{t}$ on the
reduced frequency $\tilde{\omega} = \omega/ \omega_{r}$ is depicted in Fig.~2,
curve 1, for a system of spherical nanoparticles with $H_{a} = 6400\,$Oe, $m/V
= 1400\,$G ($V$ is the nanoparticle volume), $r = 4\,$nm ($r$ is the
nanoparticle radius), $\lambda = 10^{-2}$, $h = 10\,$Oe, $\rho = -1$, and $T =
300\,$K. Note that for these parameters $a = H_{a}m/2kT \approx 29$, $\mu_{t}
|_{\tilde {\omega} = 1} \approx 0.34$, $\mu_{d} |_{\tilde{\omega} = 1} \approx
6.1 \times 10^{-3}$, and $\mu_{t}/ \mu_{d} \approx 56$.

Thus, the above results  show  that the magnetic field rotating in the plane
perpendicular to the easy axes of magnetic nanoparticles changes the
probabilities of the up and down orientations of the nanoparticle magnetic
moments. Due to thermal fluctuations, the nanoparticle system magnetizes in the
direction with larger probability. In the case of nanoparticles with large
anisotropy barrier (when $a \gg 1$) this contribution, i.e. $\mu_{t}$, to the
total magnetization $\mu$ considerably exceeds a small dynamical contribution,
i.e. $\mu_{d}$, which is induced by the rotating magnetic field at $T = 0$. We
emphasize that the magnetization of nanoparticle systems in the rotating
magnetic field arises from the different dynamical behavior of the magnetic
moments in the up and down states. In turn, the difference in the dynamics of
the magnetic moments results from the existence of a well-defined direction of
their natural precession (counterclockwise when viewed from above).

\textit{Role of dipolar interactions.}---To check the role of the magnetic
dipolar interaction, we performed a Monte Carlo simulation for two-dimensional
arrays of dipolar interacting nanoparticles, representing an important class of
patterned magnetic recording media \cite{nanostruc}. In doing so we assumed
that the centers of $N$ nanoparticles occupy the sites of a square lattice of
size $Ld \times Ld$ ($L$ is a natural number, $(L+1)^{2} = N$, $d$ is the
lattice spacing). The easy axes of the nanoparticles are perpendicular to the
lattice plane and the magnetic field rotates in this plane. In contrast to the
previous case, the dipolar magnetic field acts on each magnetic moment. This
field is changed from site to site and, due to the random motion of the
magnetic moments, fluctuates with time. For $a \gg 1$, the fluctuations of the
magnetic moments and the rates of their reorientations are small. Thus,  the
dipolar field acting on the $i$-th magnetic moment during the $l$-th step can
be approximated as $\mathbf{H}_ {i}(l) = (0,0,H_{i}(l))$, where $H_{i}(l) =
-m\sum_{j\neq i} \sigma_{j}(l)/ r_{ij}^{3}$, $\sigma_{j}(l) = +1$ or $-1$, and
$r_{ij}$ is the distance between the centers of the nanoparticles. In this
case, the magnetization of the nanoparticle system can be represented through
the step-dependent magnetization $\mu_{t}(l) = \sum_{i=1}^{N} \sigma_{i}(l)/N$
as follows:
\begin{equation}
    \langle\mu_{t}\rangle = \frac{1}{l_{2} - l_{1} + 1}\sum_{l =
    l_{1}}^{l_{2}}\mu_{t}(l).
    \label{eq:av_mu}
\end{equation}
Here, to be sure that the system reached the steady state and the averaging
procedure is correct, the conditions $l_{1} \gg 1$ and $l_{2} - l_{1} \gg 1$
are implied.

In order to apply the Monte Carlo method for calculating $\langle \mu_{t}
\rangle$, we need to evaluate for all $i$ and $l$ the probabilities
$p_{\sigma_{i}}(i,l)$ that the $i$th magnetic moment stays in the states
$\sigma_{i}(l)$. But in our case the conventional approach involving the
Boltzmann factor for the solution of this problem is not applicable because the
rotating field depends on time. Therefore, we extended the above method to the
case of dipolar interacting nanoparticles and calculated the mean times that
the $i$th magnetic moment spends in the up and down states:
\begin{equation}
    t_{\sigma_{i}} = t_{0} \frac{\exp{[a(\cos\theta_{\sigma_{i}} +
    \sigma_{i}b_{i})^{2}} - a]} {(1-b_{i}^{2})(\cos\theta_{\sigma_{i}} +
    \sigma_{i}b_{i})},
    \label{eq:as_t_sigma_i}
\end{equation}
where $b_{i} = H_{i}(l)/H_{a}$ and $\theta_{\sigma_{i}}$ is the precession
angle of the $i$th magnetic moment that is defined by Eq.~(\ref{theta_sigma})
in which $\sigma$ should be replaced by $\sigma_{i}(l)$ and $\omega_{r}$ by
$\omega_{r} + \gamma H_{i}(l)$ (for brevity, we omitted the arguments $i$ and
$l$). Next, defining $p_{\sigma_{i}}(i,l) = t_{\sigma_{i}}(i,l) / [t_{+1}(i,l)
+ t_{-1}(i,l)]$ and using the numerical procedure developed in \cite{DLT}, we
performed a Monte Carlo simulation of the magnetization $\langle \mu_{t}
\rangle$ induced by the rotating field in two-dimensional systems of dipolar
interacting nanoparticles.

In Fig.~2, curve 2, we depict the dependence of $\langle\mu_{t}\rangle$ on
$\tilde{\omega}$ for the square array of the same nanoparticles driven by the
same rotating field at $N = 10^{4}$, $d = 5r$, $\eta = 10^{-3}$, $l_{1} =
10^{3}$, and $l_{2} = 5\times 10^{3}$. Comparing the curves 1 and 2 shows that
the magnetic dipolar interaction reduces the induced magnetization, widens its
frequency dependence, and raises the resonance frequency ($\tilde {\omega}_{m}>
1$). We emphasize that although the dipolar interaction reduces the induced
magnetization, its experimental observation is still possible even in this
strongly interacting case. Note also that  $\mu _{t}^{mf}$, which we evaluated
within the mean-field approximation (see Fig.~2, curve 3), distinctly differs
from $\langle \mu_{t} \rangle$ because this approximation does not account for
the crucial feature of the dipolar interaction in these systems, i.e., its
antiferromagnetic character.

\textit{Potential applications.}---The above results evidence that the
frequency dependence of the induced magnetization is detectable and because the
magnetic resonance methods are both very accurate and sensitive, its
experimental determination thus provides valuable information about the dipolar
field distribution in such systems. In particular, the average dipolar field
acting on the resonant particles can  approximately be estimated as $\rho H_{a}
(\tilde {\omega}_{m} - 1)$. Moreover, due to the selective change of the
thermal stability of the magnetic moments, which is controlled by the
characteristics of the rotating magnetic field, it calls for potential
applications in magnetic recording technology.

\textit{Resume.}---We succeeded to show that a small dynamical magnetization of
nanoparticle systems that is induced by the circularly polarized magnetic field
is strongly enhanced by thermal fluctuations. The thermally enhanced
magnetization  exhibits as a function of the field frequency a resonant
character, possessing a well pronounced extremum. The magnetic dipolar
interaction increases the relative width of the frequency dependence of
magnetization while causing a decrease of its strength and a corresponding
shift of its maximum to higher frequencies.

S.I.D. and T.V.L. acknowledge the support of the EU through NANOSPIN Contract
No. NMP4-CT-2004-013545, S.I.D. acknowledges the support of the EU Contract No.
MIF1-CT-2005-007021, and P.H. acknowledges the support of the DFG via the SFB
486.

\newpage

\begin{figure}[h]
    \centering
    \caption{\label{fig1} Sketch of the
    precession of the up and down magnetic moments
    (arrows show the directions of their natural precession)
    at $\rho = -1$.}
\end{figure}

\begin{figure}[h]
    \centering
    \caption{\label{fig2} Plots of $\mu_{t}$ (curve 1), $\langle
    \mu_{t} \rangle$ (curve 2), and $\mu_{t}^{mf}$ (curve 3) as
    the functions of the reduced frequency $\tilde{\omega}$.}
\end{figure}

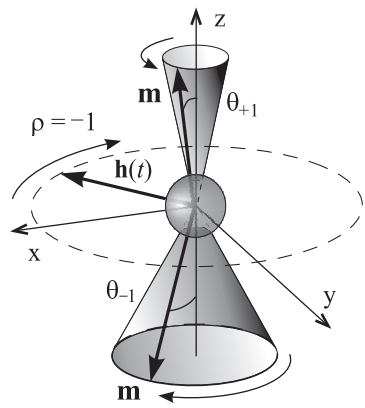
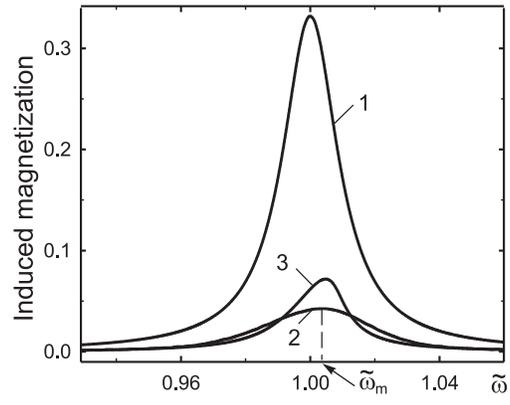
\end{document}